\documentclass[pra,aps,superscriptaddress,twocolumn]{revtex4}

\usepackage{amsmath}
\usepackage{amsfonts}
\usepackage{bm}
\usepackage{graphicx}
\usepackage{color}

\begin{document}

\title{Neural-network quantum states at finite temperature}

\author{Naoki Irikura}
\affiliation{Department of Engineering Science, University of
Electro-Communications, Tokyo 182-8585, Japan}

\author{Hiroki Saito}
\affiliation{Department of Engineering Science, University of
Electro-Communications, Tokyo 182-8585, Japan}

\date{\today}

\begin{abstract}
We propose a method to obtain the thermal-equilibrium density matrix of a
many-body quantum system using artificial neural networks.
The variational function of the many-body density matrix is represented by a
convolutional neural network with two input channels.
We first prepare an infinite-temperature state, and the temperature is
lowered by imaginary-time evolution.
We apply this method to the one-dimensional Bose-Hubbard model and compare
the results with those obtained by exact diagonalization.
\end{abstract}

\maketitle

\section{Introduction}

One of the challenging problems in physics is the determination of the
properties of quantum many-body systems.
Quantum many-body problems are difficult to solve, since the size of
the Hilbert space exponentially increases with the size of the system.
An approximate method to overcome this difficulty is the variational method,
in which the huge Hilbert space is represented by a variational wave
function with a tractable number of variational parameters.
However, the variational method relies greatly on the physical insight of
researchers to find sophisticated variational wave functions~\cite{Feynman,
BCS}.

Carleo and Troyer~\cite{Carleo} proposed the use of artificial neural
networks to represent variational wave functions for quantum many-body
states.
It is known that artificial neural networks are very flexible and can
approximate any function if the number of hidden units in the neural
networks is sufficient.
Using artificial neural networks as variational functions, therefore, we
expect that quantum many-body wave functions can be approximated
efficiently, in which the essential features of quantum many-body states are
automatically captured as variational network parameters are optimized.
This method has been applied to a variety of quantum many-body
problems, and various properties of quantum many-body states represented by
neural networks have been investigated~\cite{Deng, Chen, Cai, Gao, SaitoL,
  Nomura, Kato, Glasser, Carleo08, Czischek, Ruggeri, Saito18, Levine,
  ChooL, Liang, Luo, Lu, Choo}.

Recently, artificial neural networks were also used to represent the density
matrices of open quantum many-body systems~\cite{Torlai, Yoshioka, Nagy,
  Hartmann, Vicentini}.
A density operator $\hat\rho$ contains more information than a pure state
$|\psi\rangle$, and open quantum systems need more representation ability of
neural networks than closed quantum systems.
In Refs.~\cite{Yoshioka, Nagy, Hartmann, Vicentini}, the master equations in
the Lindblad form are solved using the variational Monte Carlo method, and
the steady states of dissipative spin systems are obtained.
The successful use of neural networks to represent density matrices opens up
the application of machine learning not only to dissipative quantum systems,
but also to finite-temperature states of quantum many-body systems.

Although the Boltzmann machine was used in the previous
studies~\cite{Torlai, Yoshioka, Nagy, Hartmann, Vicentini}, in this paper,
we use a convolutional neural network (CNN)~\cite{textbook} to represent the
density matrix of a finite-temperature state.
The CNN has been used to represent the ground states, i.e., pure states, of
quantum many-body systems~\cite{Kato,Levine,Liang,Choo}.
In the case of the pure state $|\psi \rangle$, for the base
$|\bm{n}\rangle$, the configuration of particles or spins $\bm{n}$ is input
into the CNN, and the output of the CNN gives the amplitude $\langle \bm{n}
| \psi \rangle$.
For the density matrix, in the present study, we input $\bm{n}$ and
$\bm{n}'$ into the CNN with two input channels, and the output of the CNN
gives the matrix element $\langle \bm{n} | \hat\rho | \bm{n}' \rangle$ of
the density operator $\hat\rho$.
We first prepare the density matrix at infinite temperature with $\beta =
(k_B T)^{-1} = 0$, and the imaginary-time propagator
$e^{-\Delta\beta \hat H}$ is applied to the density matrix successively
to obtain $\hat\rho = e^{-\beta \hat H}$ at each $\beta$.
A similar imaginary-time method was used to obtain the thermal equilibrium
in matrix product states~\cite{Verstraete,Zwolak,Feiguin}.
We apply our method to the Bose-Hubbard model, which describes cold bosonic
atoms in optical lattices~\cite{Jaksch}.
We calculate the finite-temperature density matrix of the Bose-Hubbard model
in one-dimensional space and compare the results with those obtained by
exact diagonalization.
We also investigate the dependence of the accuracy of our method on various
conditions, such as CNN structures.

This paper is organized as follows.
Section~\ref{s:method} explains the method,
Sec.~\ref{s:results} shows the numerical results, and 
Sec.~\ref{s:conc} provides the conclusions of the study.

\section{Method}
\label{s:method}

To demonstrate the neural-network method for obtaining the
finite-temperature density matrix, we apply it to the Bose-Hubbard model in
one-dimensional space.
The Hamiltonian is given by
\begin{equation}
\hat H = -\sum_{\langle i, j \rangle}{\hat a_i \hat a_j^\dagger}
+ \frac{U}{2}\sum_i \hat n_i (\hat n_i - 1),
\end{equation}
where $U$ is the on-site interaction energy, $\hat a_i$ is the annihilation
operator of a boson at the $i$th site, $\hat n_i = \hat a_i^\dagger \hat
a_i$ is the number operator, and $\langle i, j \rangle$ represents adjacent
sites.
The energy is normalized in such a way that the hopping coefficient becomes
unity.
Such a system can be realized by ultracold bosonic atoms loaded in an
optical lattice~\cite{Jaksch}.
We assume the periodic boundary condition, $\hat a_{M+1} = \hat a_1$, where
$M$ is the number of sites.
A pure quantum state can be expanded by the Fock-state bases
$|\bm{n}\rangle$, where $\bm{n} = (n_1, n_2, \cdots, n_M)$ represents the
number of bosons in each site.
We consider a canonical ensemble at temperature $T = (k_B \beta)^{-1}$ with
total number of bosons $N$.
The number of Fock-state bases $|\bm{n}\rangle$ satisfying $\sum_i n_i = N$
is $N_{\rm   base} = (N + M - 1)! / [N! (M-1)!]$, which increases
exponentially with $N$ and $M$.
We restrict ourselves to the case of $N = M$ in the following analysis.
In this case, at zero temperature, the system becomes a Mott insulator for
large $U$, and the system exhibits superfluidity for small $U$.
At finite temperature, the normal phase emerges~\cite{Dicker} around the
Mott insulator and superfluid regions in the phase diagram.
Since all the matrix elements $\langle \bm{n} | e^{-\epsilon\hat H}
| \bm{n}' \rangle$ for infinitesimal $\epsilon > 0$ can be taken to be real
and nonnegative without loss of generality, all the matrix elements of the
thermal density matrix $\langle \bm{n} | e^{-\beta \hat H} | \bm{n}'
\rangle$, which are decomposed to matrix products of $\langle \bm{n} |
e^{-\epsilon\hat H} | \bm{n}' \rangle$, can be taken to be real and
nonnegative.

\begin{figure}[tb]
\includegraphics[width=8cm]{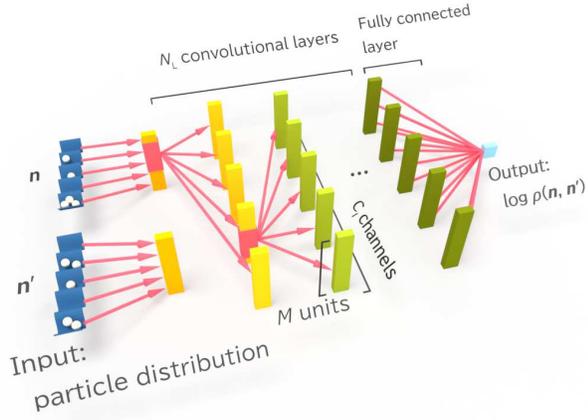}
\caption{
  (color online) Schematic illustration of the CNN to represent the density
  matrix $\rho(\bm{n}, \bm{n}') = \langle \bm{n} | \hat\rho | \bm{n}'
  \rangle$.
  The one-dimensional configurations of bosons $\bm{n}$ and $\bm{n}'$ on $M$
  sites are input into the two input channels.
  Successive $N_L$ convolutional layers are followed by a fully-connected
  layer, which gives output $u^{\rm (out)}$.
  The matrix element of the density matrix $\rho(\bm{n}, \bm{n}')$ is given
  by $e^{u^{\rm (out)}}$.
}
\label{f:schematic}
\end{figure}
We employ the CNN~\cite{textbook} to represent the density matrix $\langle
\bm{n} | \hat\rho | \bm{n}' \rangle \equiv \rho(\bm{n}, \bm{n}')$ of the
system (see Fig.~\ref{f:schematic}).
The inputs into the CNN are $\bm{n}$ and $\bm{n}'$, which we denote as
$\bm{u}^{(0)}_1$ and $\bm{u}^{(0)}_2$, respectively, i.e., the CNN has two
input channels, with each of size $M$.
The first hidden layer is calculated as
\begin{equation} \label{cnn1}
u_{m, j}^{(1)} = \sum_{k=1}^2 \sum_{p=0}^{F_1-1} W_{k, m, p}^{(1)}
u_{k, j+p}^{(0)} + b_m^{(1)},
\end{equation}
and these are propagated to the deeper layers as
\begin{equation} \label{cnn2}
u_{m, j}^{(l)} = \sum_{k=1}^{C_{l-1}} \sum_{p=0}^{F_l-1} W_{k, m, p}^{(l)}
f(u_{k, j+p}^{(l-1)}) + b_m^{(l)},
\end{equation}
where $\bm{W}_{k, m}^{(l)}$ is the one-dimensional filter with size
$F_l$,  $\bm{b}^{(l)}$ is the bias, and $C_l$ is the number of channels in
the $l$th hidden layer.
In Eqs.~(\ref{cnn1}) and (\ref{cnn2}), the subscripts $m$ and $k$ identify
the channels.
The number of units in each channel in the input and hidden layers is $M$,
i.e., $\bm{u}_{m}^{(l)} = (u_{m, 1}^{(l)}, u_{m, 2}^{(l)}, \cdots,
u_{m, M}^{(l)})$, which satisfies the periodic boundary condition,
$u_{m, M+1}^{(l)} = u_{m, 1}^{(l)}$.
We adopt the leaky ReLU~\cite{textbook} function as the activation
function $f$,
\begin{equation} \label{relu}
  f(x) =
  \begin{cases}
    x     & (x \geq 0) \\
    a x & (x < 0)
  \end{cases}
\end{equation}
with a constant $a > 0$.
After $N_L$ convolutional layers, the CNN finally gives a single output
value $u^{\rm (out)}$ through the fully-connected layer as
\begin{equation} \label{uout}
  u^{\rm (out)} = \sum_{m=1}^{C_L} \sum_{j=1}^M W^{\rm (fc)}_{m, j}
  u_{m, j}^{(L)}.
\end{equation}
The network parameters are thus the filters $\bm{W}_{k, m}^{(l)}$ and biases
$\bm{b}^{(l)}$ in the convolutional layers, and weights
$\bm{W}_m^{\rm (fc)}$ in the fully-connected layer, which are all taken to
be real, and therefore the output $u^{\rm (out)}$ is real.
In the following, we abbreviate these network parameters as $\bm{W}$.
Using the output $u^{\rm (out)}$ in Eq.~(\ref{uout}), the matrix element of
the density matrix is represented as
\begin{equation} \label{rho}
\rho(\bm{n}, \bm{n}') = e^{u^{\rm (out)}}.
\end{equation}
Although such representation of the density matrix using the CNN does
not assure its Hermiticity and positive definiteness, unlike the
Boltzmann-machine representation proposed in Ref.~\cite{Torlai}, we will see
that the Hermiticity and positive definiteness are approximately satisfied
during the imaginary-time evolution.

The imaginary-time evolution of the density matrix is realized as follows. 
Suppose that we have a CNN that represents the density matrix at inverse
temperature $\beta$, $\rho_\beta(\bm{n}, \bm{n}') = \langle \bm{n} |
e^{-\beta \hat H} | \bm{n}' \rangle$.
The value of a matrix element of the density matrix at $\beta + \Delta
\beta$ is calculated from $\rho_\beta(\bm{n}, \bm{n}')$ as
\begin{eqnarray}
& & \rho_{\beta+\Delta\beta}(\bm{n}, \bm{n}') =
\langle \bm{n} | e^{-(\beta + \Delta\beta) \hat H} | \bm{n}' \rangle
\nonumber \\
& = & \sum_{\bm{n}_1, \bm{n}_2}
\langle \bm{n} | e^{-\frac{\Delta\beta \hat H}{2}} | \bm{n}_1 \rangle
\rho_\beta(\bm{n}_1, \bm{n}_2)
\langle \bm{n}_2 | e^{-\frac{\Delta\beta \hat H}{2}} | \bm{n}' \rangle
\nonumber \\
& = & \rho_\beta(\bm{n}, \bm{n}') - \frac{\Delta\beta}{2} \sum_{\bm{n}_1}
\Bigl[
  \langle \bm{n} | \hat H  | \bm{n}_1 \rangle \rho_\beta(\bm{n}_1, \bm{n}')
\nonumber \\
& & + \rho_\beta(\bm{n}, \bm{n}_1) \langle \bm{n}_1 | \hat H  | \bm{n}'
\rangle \Bigr] + \cdots + O(\Delta\beta^{K+1}),
\label{dbeta}
\end{eqnarray}
where we expand $e^{-\Delta\beta \hat H / 2}$ with respect to $\Delta\beta$
in the last line.
We cut off the $O(\Delta\beta^{K+1})$ terms in Eq.~(\ref{dbeta}).
By this approximation, the number of terms in the last line of
Eq.~(\ref{dbeta}) is reduced to $O(M^K)$, since the number of nonzero
matrix elements $\langle \bm{n} | \hat H^K | \bm{n}' \rangle$ is $O(M^K)$.
We can thus calculate any matrix elements $\rho_{\beta+\Delta\beta}(\bm{n},
\bm{n}')$ at the inverse temperature $\beta + \Delta\beta$, when we have the
CNN that represents the density matrix at $\beta$.
We next need to construct a CNN that represents the density matrix
$\rho_{\beta+\Delta\beta}(\bm{n}, \bm{n}')$.

In general, we can optimize a CNN so as to represent a desired density
matrix $\rho_{\rm target}(\bm{n}, \bm{n}')$ by minimizing
\begin{equation}
  L = \frac{1}{2} \sum_{\bm{n}, \bm{n}'} \left[ \rho(\bm{n}, \bm{n}')
    - \rho_{\rm target}(\bm{n}, \bm{n}') \right]^2,
\end{equation}
where $\rho(\bm{n}, \bm{n}')$ is the density matrix represented by the CNN
to be optimized.
We denote the network parameters of this CNN as $\bm{W}$.
We can update $\bm{W}$ to reduce the value of $L$ using its gradient with
respect to $\bm{W}$ as
\begin{equation} \label{grad}
  \frac{\partial L}{\partial w} = \sum_{\bm{n}, \bm{n}'}
    \frac{\partial \rho(\bm{n}, \bm{n}')}{\partial w}
    \left[\rho(\bm{n}, \bm{n}') - \rho_{\rm target}(\bm{n}, \bm{n}') \right],
\end{equation}
where $w$ is one of the network parameters $\bm{W}$.
Here, instead of Eq.~(\ref{grad}), we introduce a modified gradient as
\begin{equation} \label{grad2}
\sum_{\bm{n}, \bm{n}'} P(\bm{n}, \bm{n}')
    \frac{\partial \rho(\bm{n}, \bm{n}')}{\partial w}
    \left[\rho(\bm{n}, \bm{n}') - \rho_{\rm target}(\bm{n}, \bm{n}') \right],
\end{equation}
where
\begin{equation}
P(\bm{n}, \bm{n}') = \frac{\left[ \rho(\bm{n}, \bm{n}')
    - \rho_{\rm target}(\bm{n}, \bm{n}') \right]^2}
{\sum_{\bm{n}, \bm{n}'}\left[ \rho(\bm{n}, \bm{n}')
    - \rho_{\rm target}(\bm{n}, \bm{n}') \right]^2}.
\end{equation}
Since the factor $P(\bm{n}, \bm{n}')$ emphasizes the terms with larger
deviation $\left| \rho(\bm{n}, \bm{n}') - \rho_{\rm target}(\bm{n}, \bm{n}')
\right|$, we expect more efficient convergence of $\rho(\bm{n}, \bm{n}')$
to $\rho_{\rm target}(\bm{n}, \bm{n}')$ using Eq.~(\ref{grad2}) than
Eq.~(\ref{grad}).
The form of Eq.~(\ref{grad2}) is also suitable for Metropolis sampling.
Since the summation $\sum_{\bm{n}, \bm{n}'}$ cannot be taken exactly for a
large system, we calculate the summation by the Monte Carlo method with
Metropolis sampling of $\bm{n}$ and $\bm{n}'$ with probability distribution
$P(\bm{n}, \bm{n}')$.
Using the gradient in Eq.~(\ref{grad2}), the network parameters are
updated using the Adam scheme~\cite{textbook, Kingma}, until $\rho(\bm{n},
\bm{n}')$ converges sufficiently.

We thus generate the thermal density matrix as follows.
We first prepare the initial CNN that represents the density matrix at
infinite temperature $\beta = 0$, which is used as the initial density
matrix of the imaginary-time evolution.
Such a CNN can be constructed by the method described above with 
$\rho_{\rm target}(\bm{n}, \bm{n}') = \lim_{\beta \rightarrow 0} \langle
\bm{n} | e^{-\beta \hat H} | \bm{n}' \rangle = \delta_{\bm{n}, \bm{n}'}$,
starting from random network parameters.
Next, we set the target as $\rho_{\rm target}(\bm{n}, \bm{n}') =
\rho_{\Delta\beta}(\bm{n}, \bm{n}')$, which can be calculated using
Eq.~(\ref{dbeta}).
We prepare another CNN and optimize it to represent this target, which
yields the CNN that represents $\rho_{\Delta\beta}(\bm{n}, \bm{n}')$.
Repeating this procedure, we obtain CNNs that represent
$\rho_{2\Delta\beta}$, $\rho_{3\Delta\beta}$, $\cdots$, successively.
In each step of the imaginary-time evolution, the initial values of the
network parameters for $\rho_{n \Delta\beta}$ are set to those for
$\rho_{(n-1)\Delta\beta}$ to facilitate convergence.

The expectation value of an observable $\hat A$ is written as
\begin{eqnarray} \label{expect}
  \langle \hat A \rangle
  & = & \frac{\mathrm{Tr}(\hat\rho \hat A)}{\mathrm{Tr} \hat\rho}
  = \frac{\sum_{\bm{n}, \bm{n}'} \rho(\bm{n}, \bm{n}') \langle \bm{n}' |
    \hat A | \bm{n} \rangle}
  {\sum_{\bm n} \rho(\bm n, \bm n)}
  \nonumber \\
  & = & \sum_{\bm{n}} P(\bm{n}) A(\bm{n}),
\end{eqnarray}
where $\mathrm{Tr}$ indicates trace, $P(\bm{n}) = \rho(\bm{n}, \bm{n}) /
\sum_{\bm{n}} \rho(\bm{n}, \bm{n})$ and $A(\bm{n}) = \sum_{\bm{n}'}
\rho(\bm{n}, \bm{n}') \langle \bm{n}' | \hat A | \bm{n} \rangle /
\rho(\bm{n}, \bm{n})$.
The summation in the second line of Eq.~(\ref{expect}) is calculated by the
Monte Carlo method with Metropolis sampling of $\bm{n}$ with probability
distribution $P(\bm{n})$.

\section{Results}
\label{s:results}

\begin{figure}[tb]
\includegraphics[width=8cm]{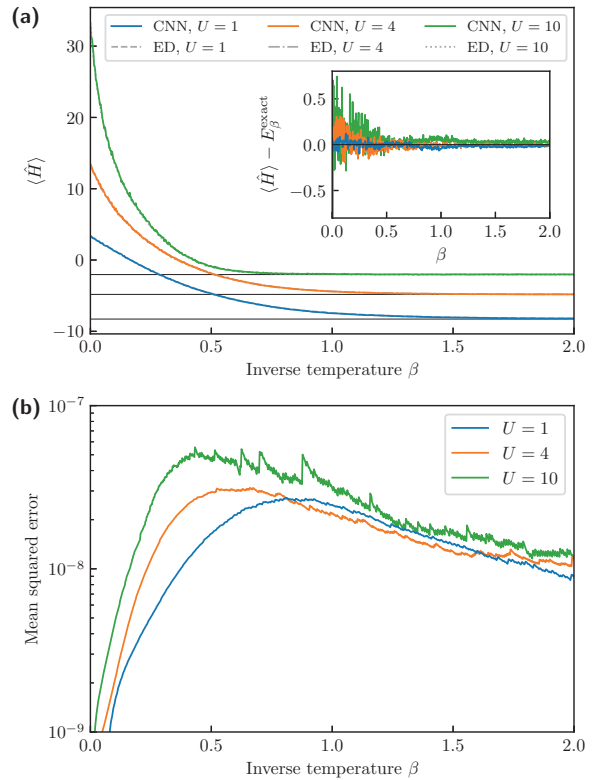}
\caption{
  (color online) Imaginary-time evolution of the density matrix represented
  by a CNN for $U = 1$, 4, and 10.
  (a) Expectation value of the Hamiltonian $\langle \hat H \rangle$.
  The lines of the energies $E_\beta^{\rm exact}$ obtained by exact
  diagonalization (ED) are also drawn, which however almost overlap with the
  lines obtained by our method and cannot be seen.
  The horizontal lines represent the exact energies of the ground states.
  The inset shows the error in the energy $\langle \hat H \rangle -
  E_\beta^{\rm exact}$.
  (b) Mean-squared error $\delta\rho$ defined in Eq.~(\ref{D}).
}
\label{f:main}
\end{figure}
We consider a system of $M = 5$ sites with $N = 5$ particles.
The CNN consists of $N_L = 4$ convolutional layers with filter size $F_1 =
F_2 = F_3 = F_4 = 5$ and $C_1 = C_2 = C_3 = C_4 = 16$ channels.
The constant in the leaky ReLU in Eq.~(\ref{relu}) is taken to be $a =
0.3$.
The imaginary-time evolution is generated with $\Delta\beta \sim
10^{-3}$, where we take the terms up to the second order of $\Delta\beta$ in
the expansion in Eq.~(\ref{dbeta}) (i.e., $K = 2$).
We take 2000 samples in the Metropolis sampling to calculate the gradient in
Eq.~(\ref{grad2}) in each Adam optimization step.
The optimization steps are performed 2000 times to obtain the next density
matrix $\rho_{\beta+\Delta\beta}$ from $\rho_\beta$ in the imaginary-time
evolution.
In order to avoid exponential growth or decay of $\langle \bm{n} |
e^{-\beta\hat H} | \bm{n}' \rangle$ in the imaginary-time evolution, we
add an appropriate constant to the Hamiltonian in each step.

Figure~\ref{f:main}(a) shows the imaginary-time evolution of the expectation
value of the energy $\langle \hat H \rangle$ obtained by our method for $U =
1$, $4$, and $10$.
In Fig.~\ref{f:main}(a), we also plot the exact energy $E_\beta^{\rm exact}$
obtained by the exact diagonalization of the Hamiltonian.
The lines of $\langle \hat H \rangle$ almost overlap with those of
$E_\beta^{\rm exact}$.
The error in the energy is $|\langle \hat H \rangle - E_\beta^{\rm exact}|
\sim 0.1$.
In Fig.~\ref{f:main}(b), we plot the mean-squared error in the matrix
elements of the density matrix, defined as~\cite{Hartmann}
\begin{equation} \label{D}
  \delta\rho = \frac{1}{N_{\rm base}^2} \sum_{\bm{n}, \bm{n}'}
  \left[ \rho_\beta(\bm{n}, \bm{n}')
    - \rho_\beta^{\rm exact}(\bm{n}, \bm{n}') \right]^2,
\end{equation}
where $\rho_\beta^{\rm exact}(\bm{n}, \bm{n}')$ is the density matrix
obtained by exact diagonalization of the Hamiltonian, and the summation is
taken over all $\bm{n}$ and $\bm{n}'$.
In calculating $\delta\rho$, the density matrix is normalized as
$\sum_{\bm{n}} \rho_\beta(\bm{n}, \bm{n}) = 1$.
The error $\delta\rho$ in Fig.~\ref{f:main}(b) is less than $10^{-7}$.
Thus, our method works well for whole temperature region and both for
superfluid and Mott insulator regimes.

In Fig.~\ref{f:main}(b), $\delta\rho$ increases for the early stage of the
imaginary-time evolution ($\beta \lesssim 0.5$), and then $\delta\rho$
decreases with $\beta$.
This is because the imaginary-time evolution $e^{-\Delta\beta \hat H / 2}
\hat\rho e^{-\Delta\beta \hat H / 2}$ eliminates excited states in
$\hat\rho$, and then also eliminates errors arising during the
imaginary-time evolution.
For $\beta \rightarrow \infty$, the density matrix converges to the ground
state, even if errors arise during the imaginary-time evolution.

\begin{figure}[tb]
\includegraphics[width=8cm]{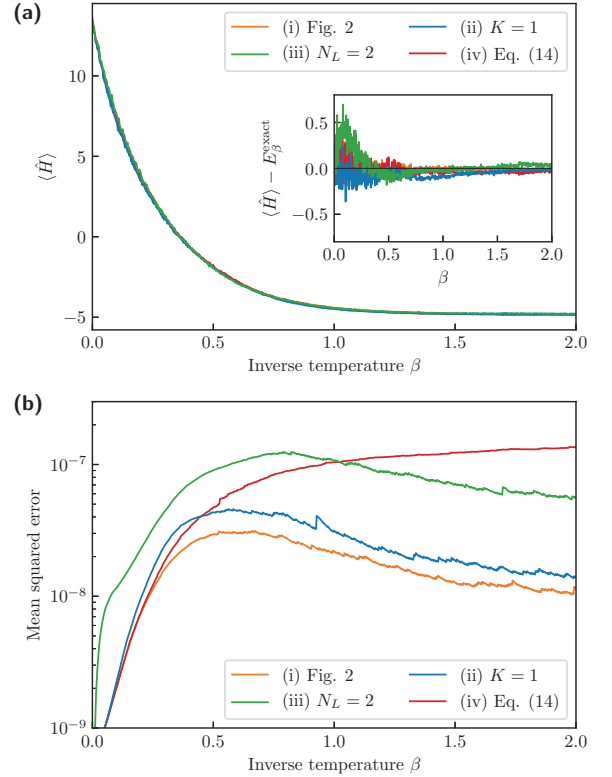}
\caption{
  (color online) Dependence of the accuracy on various conditions for $U =
  4$, where (i) the same as in Fig.~\ref{f:main}, (ii) the cutoff order in
  Eq.~(\ref{dbeta}) is reduced to $K = 1$, (iii) the number of convolutional
  layers is reduced to $N_L = 2$, and (iv) the asymmetric expansion in
  Eq.~(\ref{dbeta2}) is used with $K = 2$.
  (a) Expectation value of the Hamiltonian $\langle \hat H \rangle$.
  The inset shows the error in the energy $\langle \hat H \rangle -
  E_\beta^{\rm exact}$.
  (b) Mean-squared error $\delta\rho$ defined in Eq.~(\ref{D}).
}
\label{f:depend}
\end{figure}
The errors arise from various sources:
the cutoff in the expansion in Eq.~(\ref{dbeta}),
the representation ability of the CNN,
the statistical errors due to the Monte Carlo sampling,
and insufficient convergence in the Adam optimization.
Figure~\ref{f:depend} shows the dependence of the results on various
conditions.
We see that the errors are increased by reducing the cutoff order in
Eq.~(\ref{dbeta}) from $K = 2$ to $K = 1$ [the line (ii) in
Fig.~\ref{f:depend}] or reducing the number of convolutional layers from
$N_L = 4$ to $N_L = 2$ [the line (iii)].
We also confirmed that the accuracy is lowered by reducing the number of
samples in the Metropolis sampling or the number of iterations in the Adam
optimization (data not shown).
In Fig.~\ref{f:depend}, we also examine a different form of expansion of
$e^{-\Delta\beta \hat H}$ instead of the symmetric expansion in
Eq.~(\ref{dbeta}):
\begin{eqnarray}
\rho_{\beta+\Delta\beta}(\bm{n}, \bm{n}') & = & 
\langle \bm{n} | e^{-(\beta + \Delta\beta) \hat H} | \bm{n}' \rangle
\nonumber \\
& = & \sum_{\bm{n}''}
\langle \bm{n} | e^{-\Delta\beta \hat H} | \bm{n}'' \rangle
\langle \bm{n}'' | e^{-\beta \hat H} | \bm{n}' \rangle
\nonumber \\
& \simeq & \sum_{k=1}^{K} \frac{(-\Delta\beta)^k}{k!}
\sum_{\bm{n}''}
\langle \bm{n} | \hat{H}^k | \bm{n}'' \rangle
\rho_\beta(\bm{n}'', \bm{n}'),
\nonumber \\ \label{dbeta2}
\end{eqnarray}
where the propagator $e^{-\Delta\beta \hat H}$ always operates from the
left-hand side of $\rho_\beta$.
The line (iv) in Fig.~\ref{f:depend} shows the result using this asymmetric
expansion with $K = 2$.
The mean-squared error $\delta\rho$ monotonically increases for
Eq.~(\ref{dbeta2}).
This is because the operator $e^{-\Delta\beta \hat H}$ in the asymmetric
form in Eq.~(\ref{dbeta2}) only eliminates excited states in the ket vectors
in the density operator, and therefore, once errors arise in the bra vectors
during the imaginary-time evolution, the errors remain for $\beta
\rightarrow \infty$.

\section{Conclusions}
\label{s:conc}

We proposed a method to represent a many-body density matrix using a
convolutional neural network (CNN), where the particle configurations
$\bm{n}$ and $\bm{n}'$ are input into the CNN to produce the value of
$\langle \bm{n} | \hat \rho | \bm{n}' \rangle$.
We also proposed a method to obtain the density matrix at finite temperature
through the imaginary-time evolution of the density matrix represented by
the CNN.
We applied our method to the one-dimensional Bose-Hubbard model, and
demonstrated the imaginary-time evolution, which showed that the
finite-temperature density matrix obtained by our method agrees well with
that obtained by exact diagonalization of the Hamiltonian.
We have also investigated the dependence of the accuracy on different
conditions.

Neural network quantum states are also efficient for representing many-body
states of fermions~\cite{Nomura}, and therefore we expect that our method
can also be applied to systems with negative-sign problems, which will
be complementary to the quantum Monte Carlo method to investigate
finite-temperature properties.
The present method can also be extended to higher spatial dimensions in a
straightforward manner.

\begin{acknowledgments}
This work was supported by JSPS KAKENHI Grant Numbers JP17K05595 and
JP17K05596.
\end{acknowledgments}

\end{document}